\documentclass[12pt]{iopart}

\usepackage{graphicx}
\usepackage{subcaption}
\usepackage{etoolbox}
\usepackage{xcolor}

\begin{document}

%\maketitle

\title{Structure and Stability of the Indian Power Transmission Network}

\author{Vedang Tamhane$^1$, G Ambika$^{1,2}$}

\address{$^1$Indian Institute of Science Education and Research (IISER) Tirupati, Tirupati-517507, India}

\address{$^2$Indian Institute of Science Education and Research (IISER) TVM, Thiruvananthapuram-695551, India}

\ead{g.ambika@iisertvm.ac.in}
\vspace{10pt}

\date{\today}

\begin{abstract}
We present the study on the Indian power transmission network using the framework of a complex network and quantify its structural properties. For this, we build the network structure underlying the Indian power grid, using two of its most prevalent power lines. We construct an equivalent model of an exponential network and study {its structural changes with changes in two parameters related to redundancy and dead-ends}. Then we analyze its stability against cascading failures by varying these two parameters using the link failure model. This helps to gain insight into the relation of network topology to its stability, and {indicates how the optimum choice of these parameters can result in a power grid structure with minimum failed links. We apply the same model to study the robustness of the Indian power grid against such failures. In this case, we find that when a link connected to a generator fails, it results in a cascade that spreads in the grid until it is split into two separate stable clusters of generators and consumers, with over one-third of its nodes nonfunctional.} 

\end{abstract}
 
%\noindent{\it Keywords: Complex network; Indian power grid; cascading failures; structural stability}
%
% Uncomment for Submitted to journal-title message
%\maketitle

%-----------------------------------------------------
%----------------    Section I    --------------------
%-----------------------------------------------------
\section{\label{sec:Intro}Introduction}

Power grids are one of the largest engineering systems among the real-world complex networks. It has two types of nodes on it, power generators and consumers, connected with multiple types of connections, which makes it a complex network\cite{pagani2013power}. The efficient functioning of power grids requires synchronization in phase and frequency among all the nodes, which is studied using phase oscillators as nodal dynamics\cite{grzybowski}. However, in spite of the efforts of power engineers to maintain synchrony in such power networks, failures do occur, causing undesirable disruptions to power transmission.
The failure of a few nodes or links can disturb the synchrony, leading to disruption of the distribution of power and sometimes overloading. 
Consequently, the system is disturbed and cannot fulfill the power requirement of the loads depending on it{\cite{ZhengGao}}. In such cases, the power is supplied to demanding nodes via other paths, which causes overload in the remote nodes as well, resulting in sequential tripping, known as cascading failure{\cite{Scire}}. This cascading failure is the process that leads to undesirable blackouts that can spread in the entire power grid system{\cite{Valdez}}. The blackouts of US-Canada in 2003, China in 2005\cite{CFanalysis}, Indonesia in 2005\cite{nurdin2019indo} and India in 2012\cite{IndPG}, etc. are all realistic occurrences of the same. Hence, studies on strategies to contain the failures locally to avoid blackouts are highly relevant.

Due to the dynamic nature of its complex connections, control of power systems to optimize their performance is a challenge, requiring detailed knowledge of structural and dynamic stability. Several blackouts around the world in recent history have been investigated for their causes, and models to analyze cascading failure were proposed to arrive at emergency measures for prevention\cite{CFanalysis}. Recent studies indicate that the local patterns in the network topology influence the ability of the power grid to withstand cascading failures \cite{Deadend}. In this, using the concept of basin stability, it is shown that dead-ends diminish the stability in the artificially generated power grid model. 
The reported research on cascading failures in \cite{Valdez} explains how networks can evolve towards optimal topologies that attenuate the cascade. 
{The study by Arianos et al. \cite{arianos2009power} analyzes malicious and accidental blackouts in the power grids and their tolerant nature using complex network theory. They introduce a new parameter, net ability, to quantify the performance of power grids. 
In \cite{Schafer}, a forecasting method is proposed to help identify critical transmission lines and components in the European power grid network. 
In \cite{Hao2020} and \cite{Jia2020}, the harmonic closeness approach is used to define load on the link, and they obtain values of a tunable parameter that can be used in distributing load among links of the network. 
A study on cascading failures is conducted on weighted networks in which a local weighted flow redistribution rule is implemented \cite{Chen2008}.}

The Indian power grid is managed by Power System Operation Corporation Ltd. (POSOCO)(https://posoco.in/)\cite{posoco}. It is the single large grid that keeps the entire nation under electrification. It is composed of nearly 1575 important grid elements, which include power generation plants, distribution substations and consumers, connected via around 2200 connections. These connections are added in the cost-minimizing and simplest way of the tree-like scheme but carry different levels of electric power. The Indian power grid underwent one of its major blackouts in recent history in 2012, affecting three of its five regional grids. During maintenance, the outage in the Bina-Gwalior 400kV line led the entire grid to risk by overloading one of the North region grid stations beyond its safe limits. This event caused the cascade to spread in the East and the North-East regional grids.

In spite of being one of the most extensive grids in the world, the Indian power grid remains understudied. A complex network approach reported to investigate the blackout of 2012 in the Indian power grid\cite{IndPG} uses a practical model considering the active and reactive power loads with the locally preferential load redistribution rule.
In \cite{Panigrahi}, the structural vulnerability of the Indian power grid and random attacks on the same are discussed. 
The other studies reported include the topology and network measures of the power transmission network of Odisha\cite{dasOd} 
and West Bengal power transmission system by H Das et al. \cite{dasWB}. 
While the work of H Das et al. covers only a particular region, that by Zhang et al.\cite{IndPG} dates back to the last decade. So, we find more detailed studies of its present structure, are required to understand the subtle features underlying the whole Indian power grid.

{In this context, we present a detailed study of the Indian power grid using a complex network approach to understand its topology and compute its network measures. We first construct the Indian power grid network using the data of July 2021 made publicly available by POSOCO(https://posoco.in)and establish the exponential nature of its degree distribution. We compare its network measures with those of other well-studied power grids and standard complex networks. Using a synthetically generated network, equivalent to the Indian Power Grid, we study how the presence of dead-ends and redundancy affect the structure and topology of the network. The study of cascading failures is then carried out using the link failure model that helps to understand the dependence of its severity on the network's topology. Specifically, we study the roles of redundancy and dead-ends in the spread of cascading failures considering two types of initial failures. Then we apply the same method to the actual Indian power grid structure and analyze the extent of failed links and spread of cascades for the 400kV and the 200kV grids.

The paper is organized as follows. We present the construction of the Indian power grid from data and its structural characteristics in Section 2.  Next, we provide details on the construction of a synthetic exponential power grid and its dependence on features like redundancy and dead ends(Section 3).  In Section 4, we present the link-failure model for the spread of cascade on the constructed network.  This is then applied to the Indian Power grid in Section 5, and we end with the summary and conclusion in Section 6.}

%-----------------------------------------------------
%---------------    Section II    --------------------
%-----------------------------------------------------
\section{Structure of the Indian power grid}

The Indian power grid system is divided into five major regional grids, namely western, eastern, southern, northern and north-eastern regional grids. They are interconnected to form a single large grid with power lines of 66kV, 100kV, 110kV, 132kV, 220kV, 400kV, 765kV and HVDC(high voltage direct current) for transmitting the electric power among the different stations. The 220kV and 400kV lines are the most abundant transmission lines in the power grid of India that connect important grid elements.  

\begin{figure}[]
\centering
\includegraphics[scale=0.5]{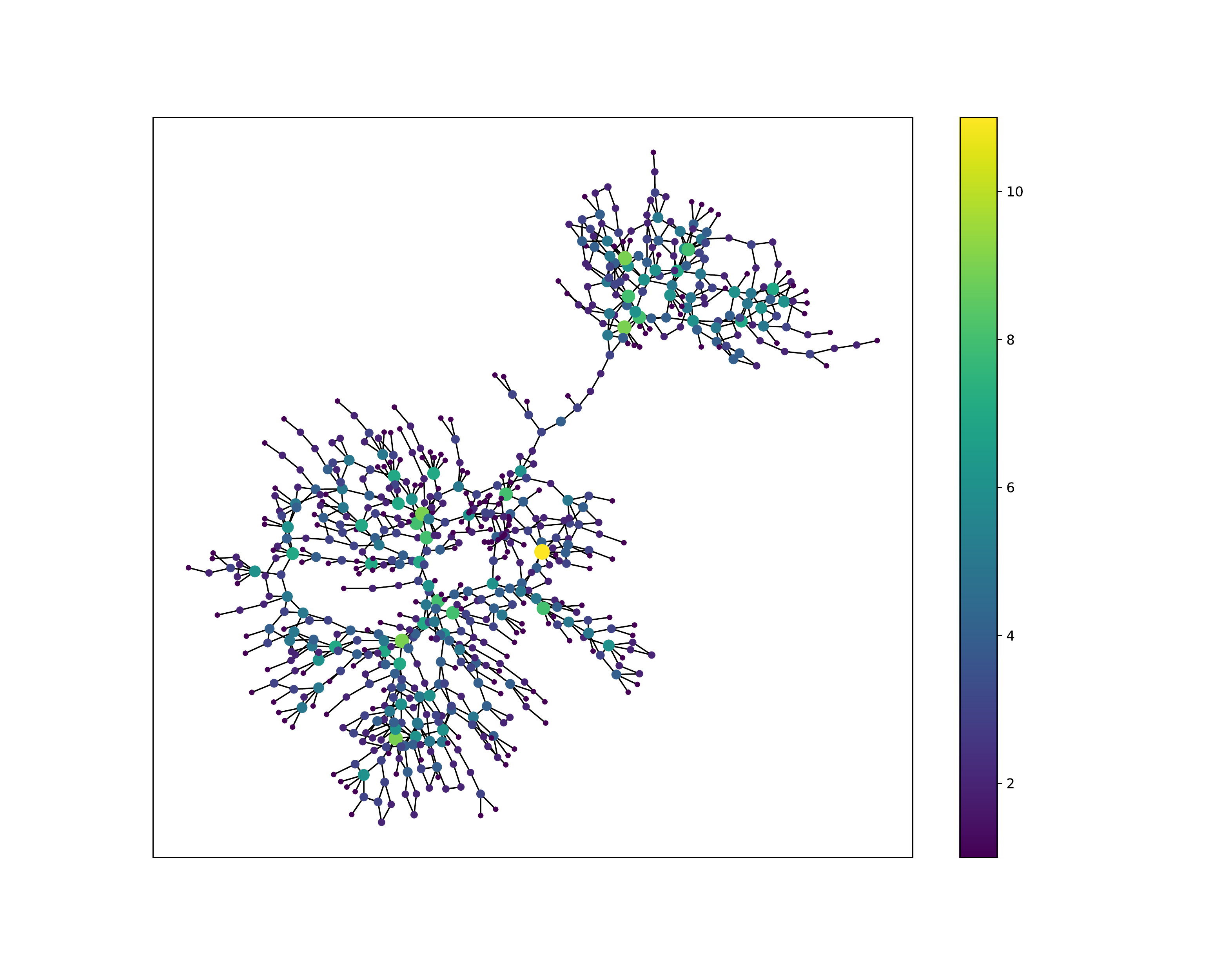}
\caption{The Indian power transmission network with 400kV lines as links. {The size of the node is proportional to its degree, while its color indicates its value.}}
\label{Fig:ntw_400}
\end{figure}

To construct the Indian power transmission network, we refer to the lists of important grid elements that are made publicly accessible by the Regional Load Dispatch Centers of POSOCO(https://posoco.i). {Since the data of the Northeastern power grid is not available, we use the maps provided by the Central Electricity Authority(CEA)(https://cea.nic.in/) that administer the working of the grid elements in this region.} The connections are then added to the edge list using the breadth-first search algorithm\cite{bfs} to get the adjacency matrix of the network.

\begin{figure}[!ht]
\centering
\includegraphics[scale=0.5]{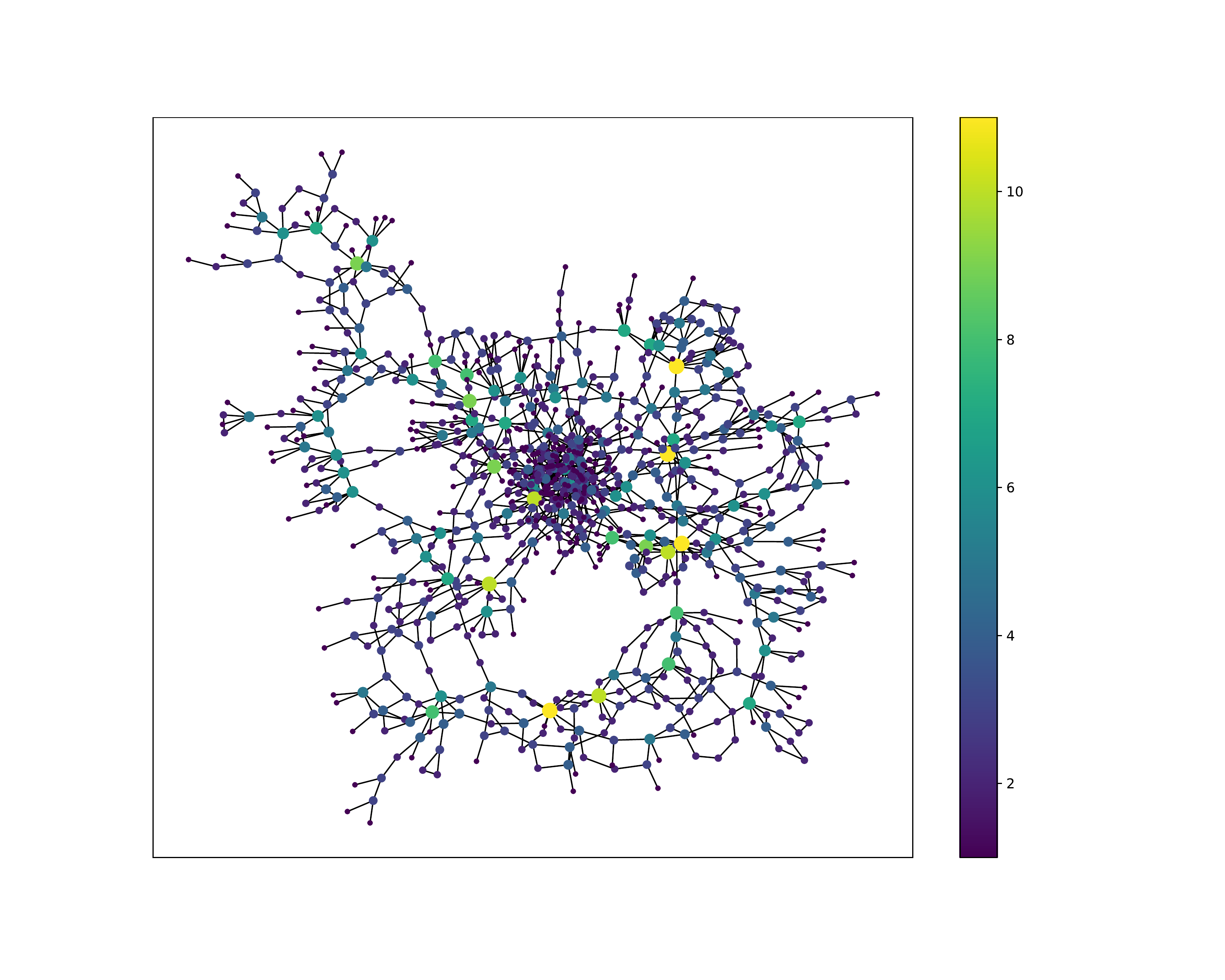}
\caption{The Indian power transmission network with 220kV lines as links. {The size of each node is proportional to its degree, while its color indicates its value.}}
\label{Fig:ntw_220}
\end{figure}

\begin{figure}[!ht]
\begin{center}
\includegraphics[scale=0.34]{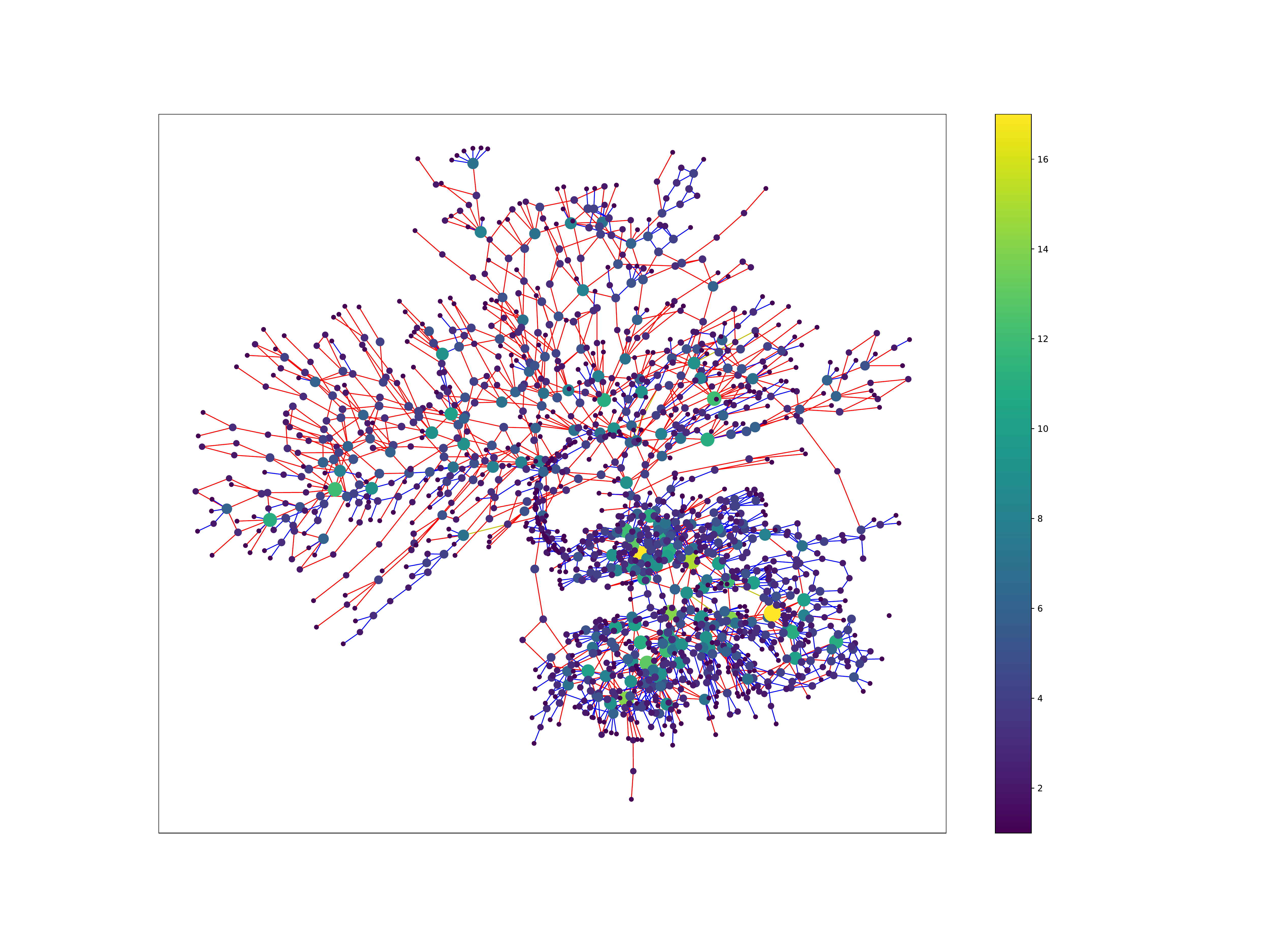}
\caption{The Indian power transmission network with both 220kV and 400kV lines included as links. The links in blue represent 220kV lines, those in red represent 400kV lines, and the yellow ones represent connections with both 220kV and 400kV lines. {The size and color of the node relate to their degree as in earlier figures.}}
\label{Fig:ntw_all}
\end{center}
\end{figure}

The diversity of the connections in the Indian power grid is such that a particular kV line does not connect all the nodes. We observe that there are small clusters in the network that are either connected by 132kV lines and lower and 765kV lines or groups of small-scale power generation plants and consumers. But 220kV and 400kV transmission lines form the largest clusters covering the entire network. The structures of the networks thus constructed with 400kV lines and 220kV lines as links(excluding a few isolated small clusters) are presented in Figures \ref{Fig:ntw_400} and \ref{Fig:ntw_220} respectively, while the network including both the lines is given in Figure \ref{Fig:ntw_all}. We use the standard methods using adjacency matrix \cite{newman} to compute the relevant measures like degree distribution, average degree, Link Density($LD$), Clustering Coefficient($CC$), Average Shortest Path Length($ASPL$) and {General Efficiency($GE$)} for a complete characterization of its topology.

\begin{figure}[!ht]
\centering
\includegraphics[scale=0.83]{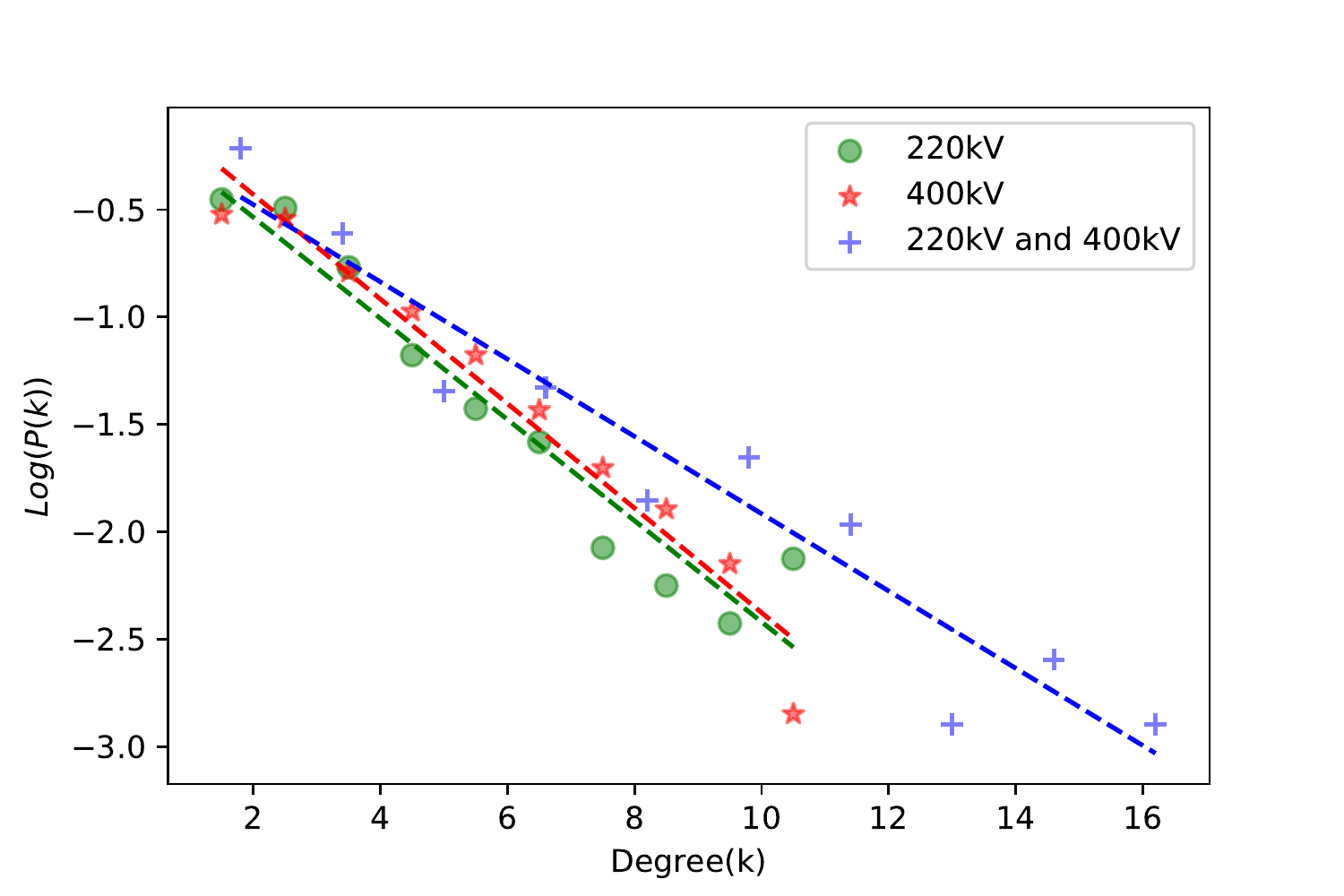}
\caption{{Semi-logarithmic plot of degree distributions of the Indian Power Transmission networks including both 220kV and 400kV transmission lines and separate networks. The linear fits indicating their exponential nature are shown using dotted lines in the same color.}}
\label{Fig:DD_all}

\end{figure}

{The degree distributions $P(k)$ of the Indian Power Transmission network, including both 220kV and 400kV transmission lines and that of the separate networks, are then computed. We plot $Log(P(k))$ for these networks in Figure \ref{Fig:DD_all}. Its linearly decreasing nature indicates the exponential nature of the network.} Therefore, we can write,

\begin{equation}
P(k) \approx e^{-\frac{k}{\gamma}}
\label{eq:ExpDD}
\end{equation}
where $k$ is the degree and $\gamma$ is the scaling index. {The plots are linearly fitted as shown in Figure \ref{Fig:DD_all} with dotted lines which gives $\gamma=1.78$ for the full network. When estimated separately, $\gamma$ comes out as 2.3 for 220 kV and 2.5 for 400 kV networks. }

We compare $\gamma$ and other topological measures like $CC$, $LD$, $\bar{k}$ and $ASPL$ of the Indian power grid with those of two other well-studied power grids, the Italian power grid and Western US power grid in Table \ref{Table:alpha}. {In addition, we include the Global Efficiency value of the largest connected component of the Indian power grids.} We also present the measures of other standard network topologies like scale-free, small-world and random networks of the same sizes(N = 1575) for comparison. 
For this, the standard networks are generated using, 
(i) Barabási-Albert model with the initial number of nodes m = 2 for a scale-free network, 
(ii) Watts-Strogatz model starting with 2 nearest neighbors and probability of rewiring p = 0.2 for a small-world and 
(iii) Erdös-Rényi graph with the probability of connection p = 0.005 for a random network\cite{newman}.
We note, compared to other standard networks with similar link density, the Indian power grid has higher $CC$ and $ASPL$.

\begin{table}[!ht]
\begin{center}
\begin{tabular}{|l c c c c c c c|}

\hline
\hline
&&&&&&&\\
{\bf Network} & {\bf $N$} & {\bf $CC$} & {\bf $GE$} &  {\bf $LD$} & {\bf $\bar{k}$} & {\bf $ASPL$} & \textbf{$\gamma$}\\
\hline
{\bf Indian Power Grid }&&&&&&& \\
220kV & 645 & 0.10 & 0.101 & 0.0022 & 2.33 & 13.00 & 2.3\\

400kV & 665 & 0.11 & 0.102 & 0.0038 & 2.66 & 14.45 & 2.5\\

220kV+400KV  & 1575 & 0.12 & 0.109 & 0.0018& 2.76 & 11.33 & 1.78\\

Italian power grid\cite{ItPG} & 341 & 0.06 & - & - & 2.46 & 16.18 & 1.8  \\

Western US PG\cite{ERGM}   & 6594 & 0.08 & - & - & 2.67 & 18.99 & 2.12 \\

&&&&&&& \\
Scale free  & 1575 & 0.0267 & - & 0.0025 & 3.99 & 4.22 & NA\\

Small world & 1575 & 0.0005 & - & 0.0015 & 2.36 & 13.43 & NA\\

Random & 1575 & 0.0045 & - & 0.0025 & 7.79 & 4.22 & NA\\
\hline
\hline
\end{tabular}
\end{center}
\caption{Characteristic measures of The Indian power grid networks compared with other power grids and standard networks.}
\label{Table:alpha}
\end{table}

\section{{Construction and Characteristics of Equivalent Exponential Network}}

We construct a synthetic exponential network that is topologically equivalent to the Indian power grid using the method described in \cite{ERGM}. This requires defining the parameters, total number of nodes $N$, number of initial nodes $N_0$, probability of connections $p$ and $q$, a measure of redundancy $r$ and measure of dead-ends $s$. 

{Following the method of construction in \cite{ERGM}, we start with a minimum spanning tree using Prim’s algorithm. Thus the network is initially coarse and tree-like that covers the spatial extents with minimal costs. During the growth, links are added to optimize the heuristic function $f(i,k,G)$, estimating the trade-off between cost and redundancy.

\begin{equation}
    f(i,j,G) = \frac{(d_G(i,j)+1)^r}{d_{spatial}(i,j)}
    \label{eq:HTF}
\end{equation}

\noindent where $i$ and $j$ are any two nodes of the network, $d_G(i,j)$ is the shortest path length between the nodes $i$, and $j$. $d_{spatial}(i,j)$ is the spatial distance between the nodes $i$ and $j$.

Based on a random number $S \in [0,1]$ generated, we construct the network as follows. If $S>s$, a node $i$ is chosen from the set of geometric coordinates, and a link is added to the node $j$ in the existing network for which $d_{spatial}(i,j)$ is minimum. Then with probability $p$, a node k is identified such that $f(i,k,G)$ is maximum and a link $i-k$ is added. With probability $q$, another pair of unconnected nodes $i^\prime$ and $j^\prime$ is chosen for which $f(i^\prime,j^\prime,G)$ is maximum and link $i^\prime-j^\prime$ is added. Hence topology changes from cost-effective to redundant depending on the value of $r$.

If $S<s$, an existing link $i-j$ is removed. A new node $h$ is added at the midpoint of $i$ and $j$, with connections, $i-h$ and $j-h$. This node gets degree k=2 while other nodes retain their degrees. Moreover, this can form ring or chain-like structures in the network, which results in less number of dead ends in the network. We repeat the above steps till we reach the total number of nodes in the networks as the chosen $N$.}

For the Indian power grid, we estimate the parameters $r=0.1$ and $s=0.3$ for the 400kV grid of size 665 and $r=0.15$ and $s=0.38$ for the 220kV grid of size 645. So, to construct a topologically equivalent network, we choose $N=655$, $N_0=100$, $r=0.1$ and $s=0.3$ and optimize probabilities as $p=0.3$ and $q=0.2$ to get $\gamma=2.4$, close to the values of these power grids. 
We then construct networks with $(N, N_0, p, q) = (655, 100, 0.3, 0.2)$ and vary the parameters $r \in [0,1.2]$ and $s \in [0,0.5]$ to study how they influence the topological features of the network. (For values of $r>1.2$ and $s>0.5$, we observe that the networks lose their characteristic exponential degree distribution). 

\begin{figure*}[ht]
\centering
\begin{subfigure}{0.5\textwidth}
\centering\includegraphics[scale=0.57]{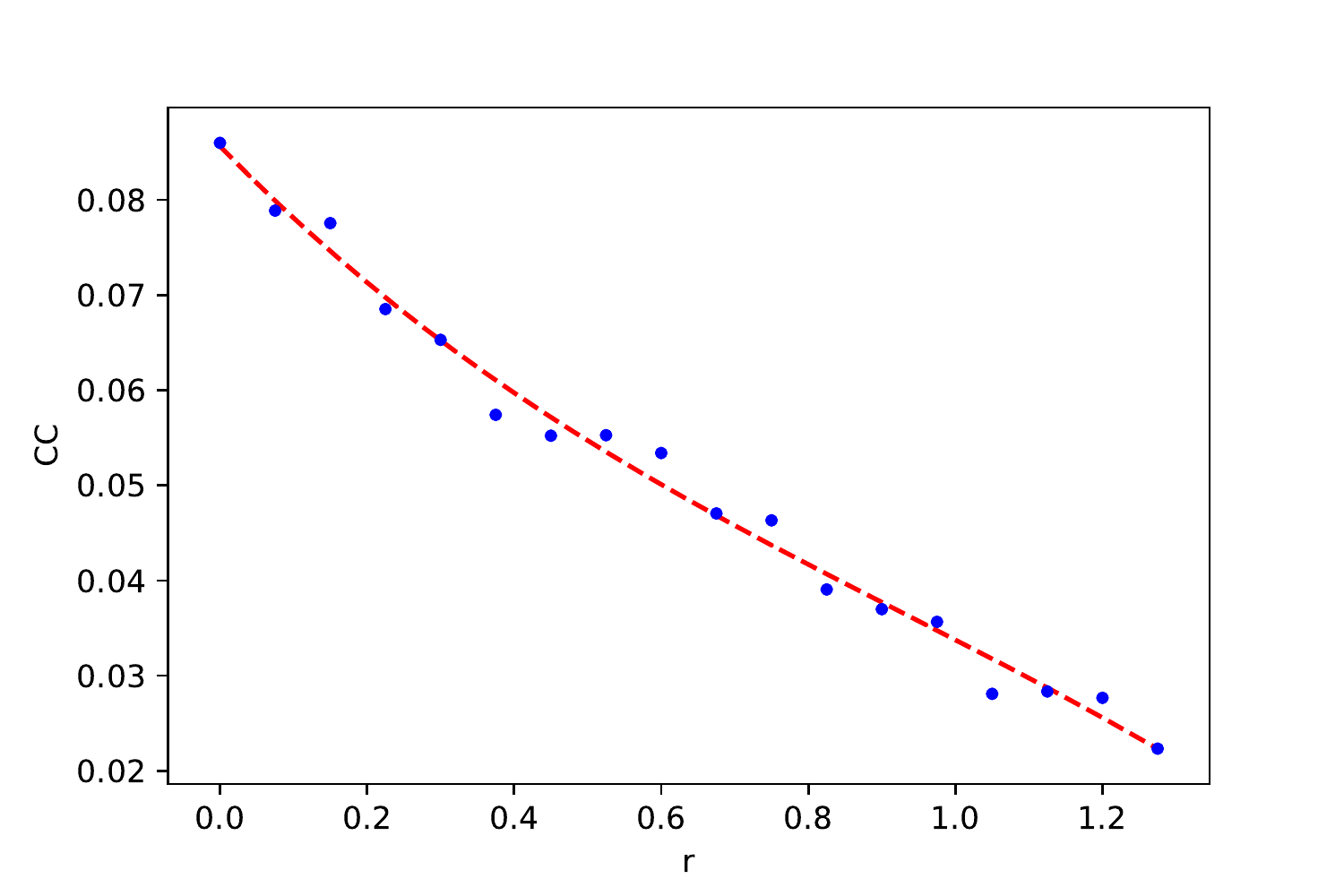}
\caption{}
\label{Fig:CCvsr}
\end{subfigure}%
~
\begin{subfigure}{0.5\textwidth}
\centering
\includegraphics[scale=0.57]{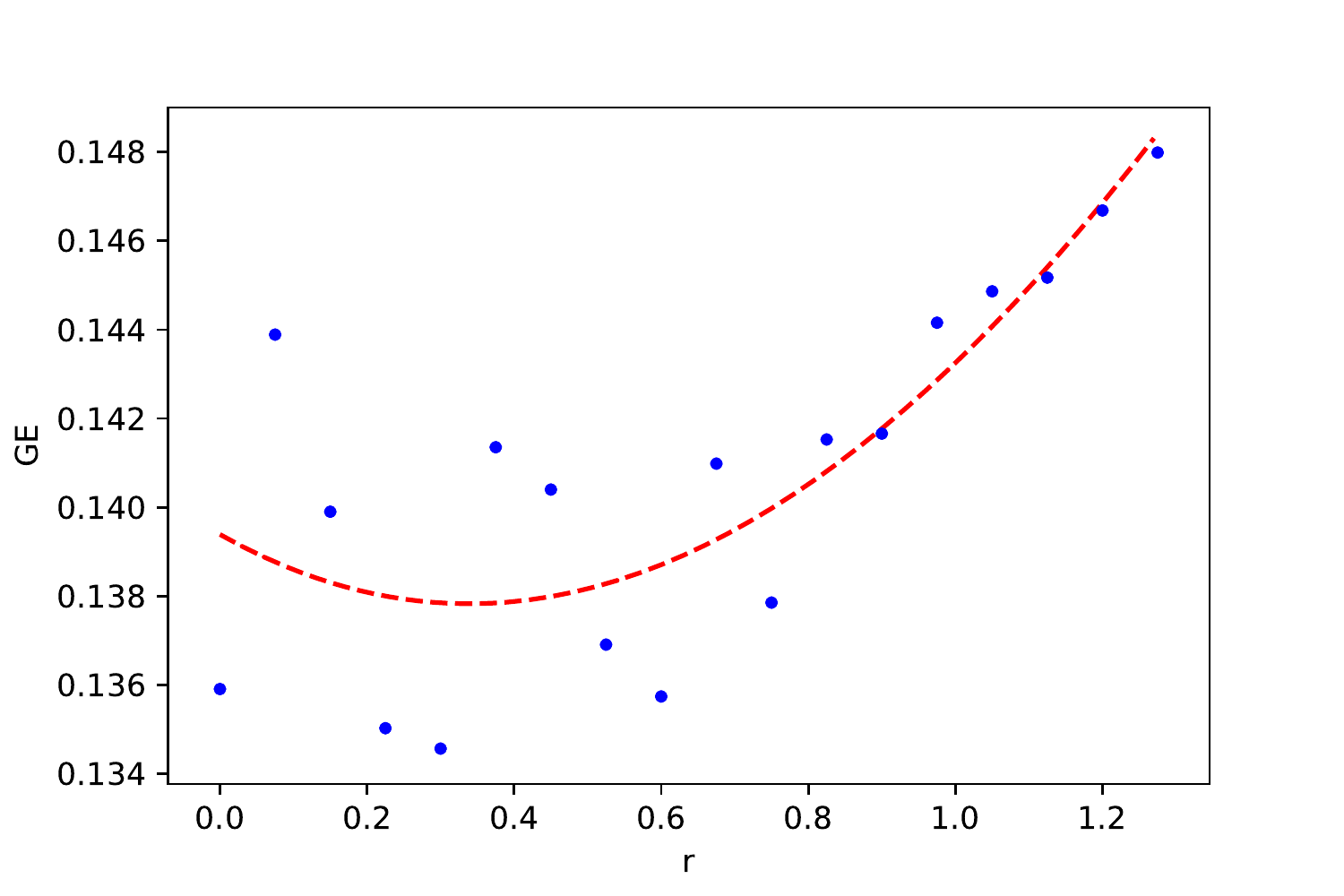}
\caption{}
\label{Fig:LDvsr}
\end{subfigure}%

\begin{subfigure}{0.5\textwidth}
\centering
\includegraphics[scale=0.57]{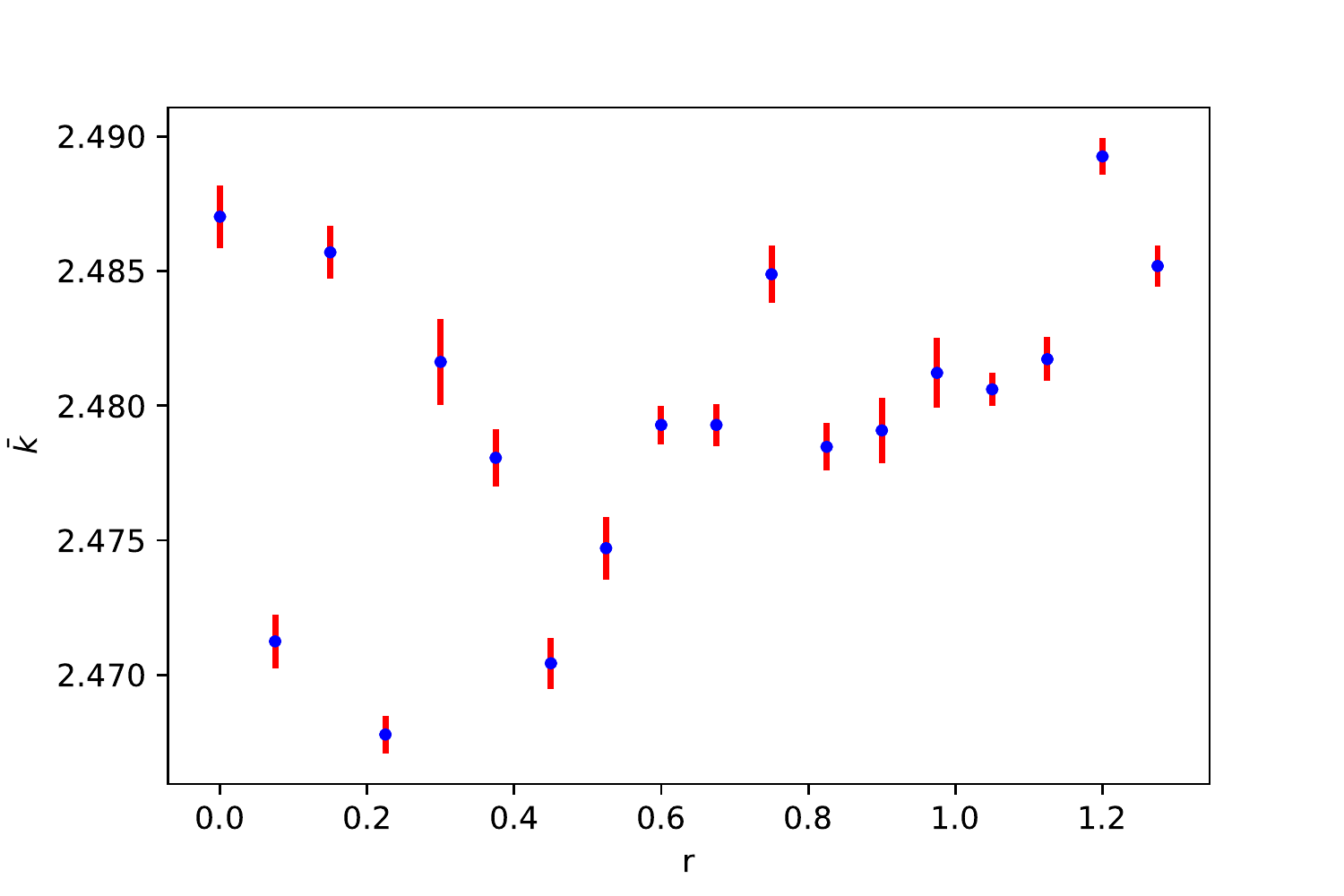}
\caption{}
\label{Fig:Kvsr}
\end{subfigure}%
\caption{{Variation of (a)Clustering Coefficient($CC$) (b)Global Efficiency ($GE$) (c)average degree($\bar{k}$) as redundancy parameter $r$ is varied keeping dead-ends parameter at $s=0.3$. The average values from 30 realizations of the constructed exponential networks are shown in each case, with their standard deviations as error bars(in red).}}
\label{Fig:VSr}
\end{figure*}

\begin{figure*}[ht]
\centering
\begin{subfigure}{0.5\textwidth}
\centering
\includegraphics[scale=0.57]{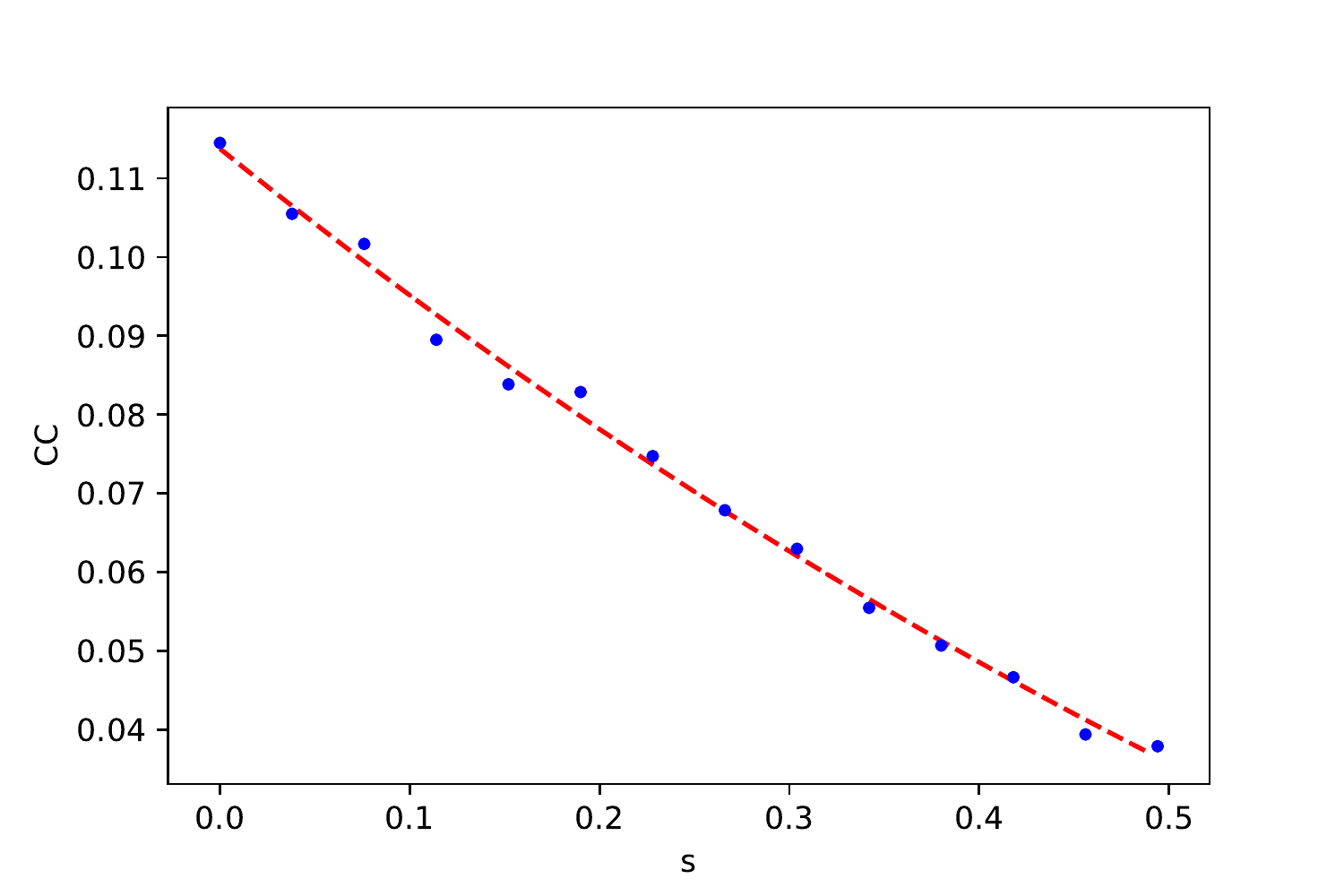}
\caption{}
\label{Fig:CCvss}
\end{subfigure}%
~
\begin{subfigure}{0.5\textwidth}
\centering
\includegraphics[scale=0.57]{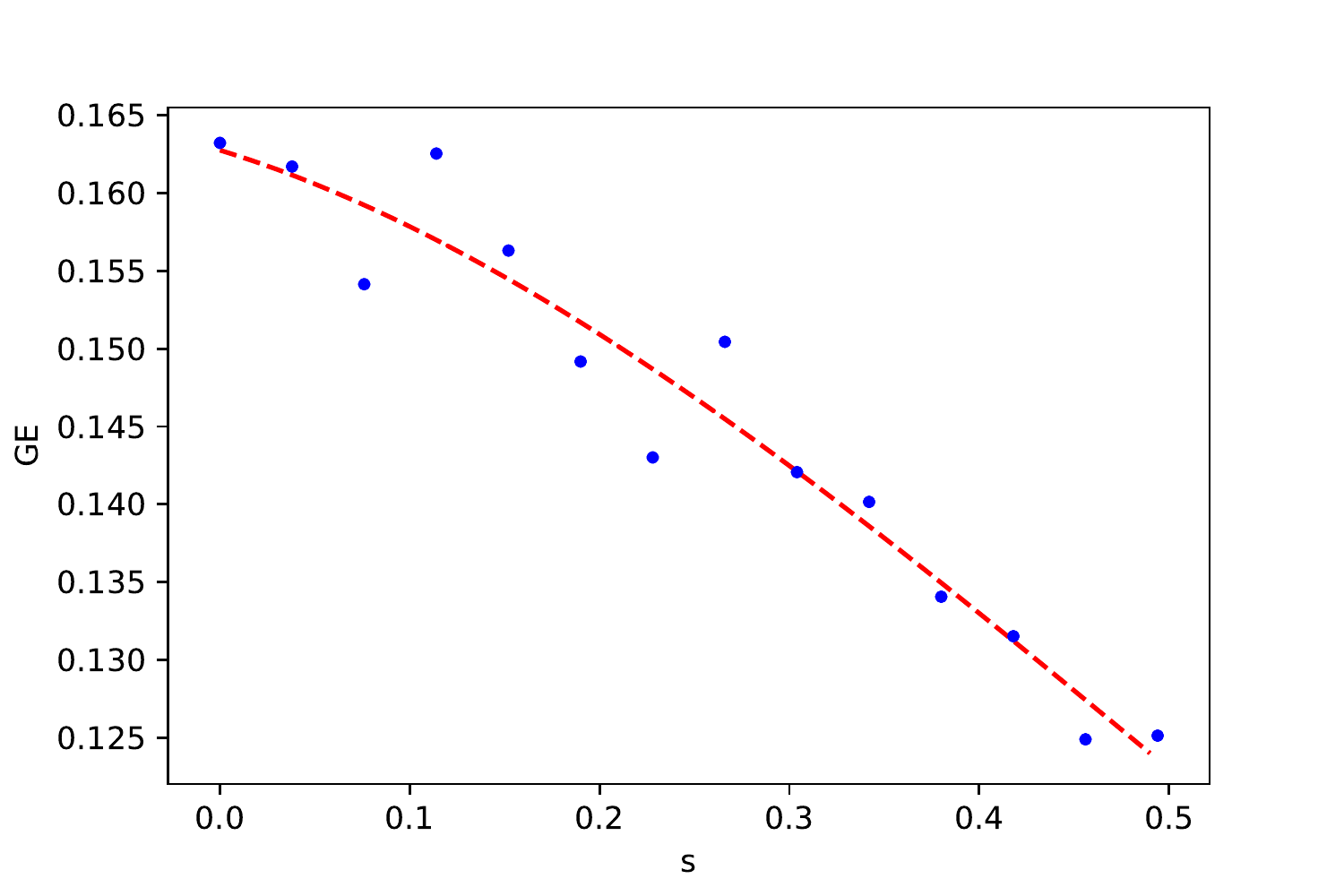}
\caption{}
\label{Fig:LDvss}
\end{subfigure}%

\begin{subfigure}{0.5\textwidth}
\centering
\includegraphics[scale=0.57]{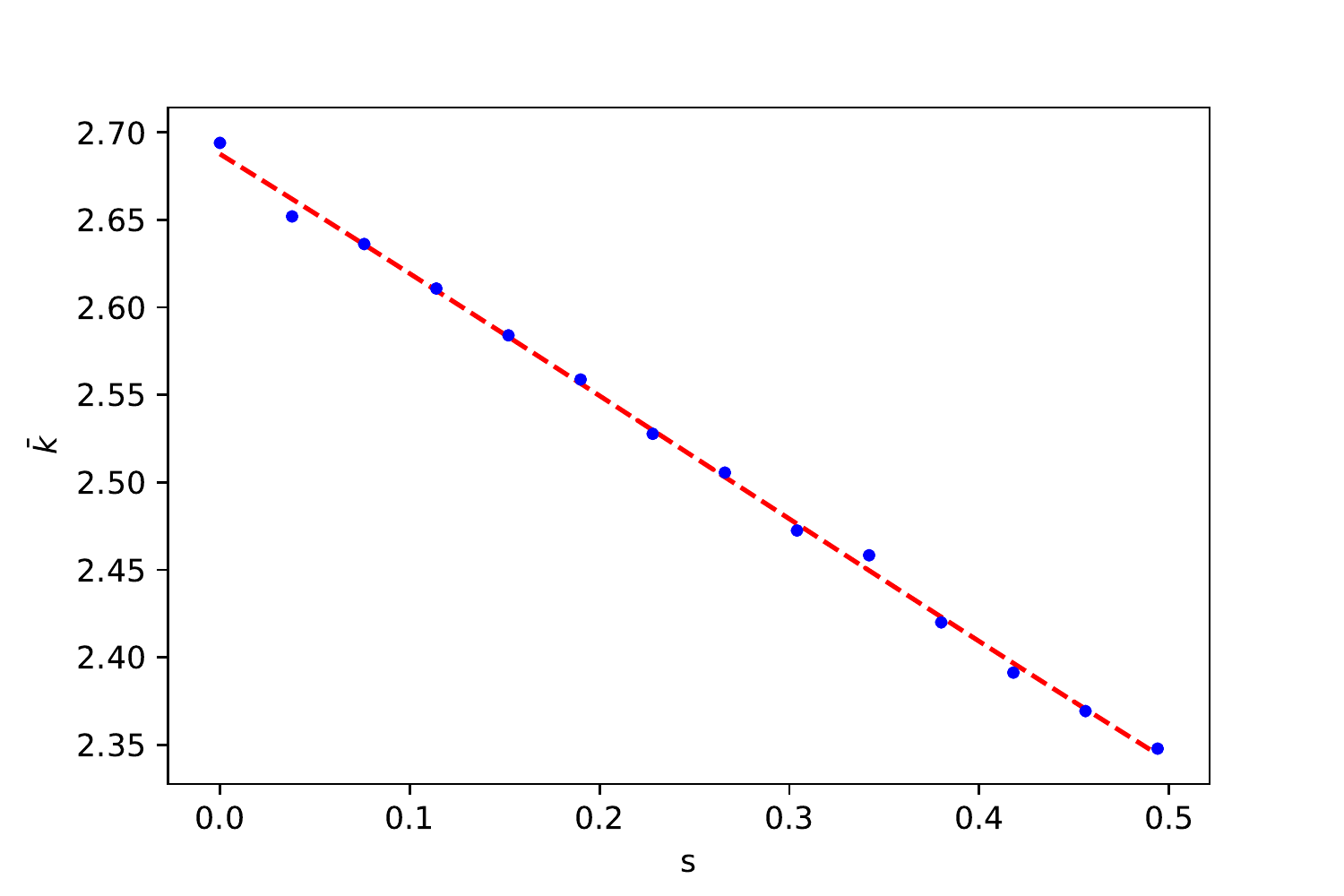}
\caption{}
\label{Fig:Kvss}
\end{subfigure}%
\caption{{Variation of (a)Clustering Coefficient($CC$) (b)Global Efficiency ($GE$) (c)average degree($\bar{k}$)as a function of $s$ for $r=0.1$. The average values are plotted from 30 realizations of the constructed exponential networks. The error bars in this case are not visible as they are very small.}}
\label{Fig:VSs}
\end{figure*}

{In Figure \ref{Fig:VSr}, we show the changes in $CC$, $GE$ and $\bar{k}$ with redundancy parameter $r$, keeping $s = 0.3$. The dotted lines indicate their trends.
We observe that as $r$ increases, $CC$ decreases while $GE$ shows an increasing trend.
By the method of construction, a new node is always linked to two more, giving equal weight to spatial and network distance. Also, in every step, two further nodes become directly linked. This produces more long-range connections in the structure. 
Increasing the parameter $r$ increases such distant redundant connections, and the local clustering coefficient, and therefore $CC$, decreases with $r$. At the same time, the average shortest path between two distant nodes decreases. This results in $GE$ increasing with $r$, while the average degree remains more or less unaffected. 

Figure \ref{Fig:VSs} shows the changes in $CC$, $GE$ and $\bar{k}$ with dead-ends parameter $s$, keeping $r = 0.1$. In this case, we find that these measures show decreasing trends.
Since increasing $s$ can form ring or chain-like structures in the network, it results in a larger average shortest path of the network with less number of triangular connections. So, $CC$ and $GE$ decreases with increase in $s$. As explained in \cite{ERGM}, the average degree depends on $s$ in such a way that it will decrease as $s$ increases.}

%-----------------------------------------------------
%---------------    Section III    -------------------
%-----------------------------------------------------
\section{Model for Cascading Failure}

The most relevant issue in the context of a power grid is its stability against local or global perturbations and natural or man-made failures. 
Most often, the cascading failure, that can lead to blackouts can be local or system-wide, depending on their origin and the topology of the network. 
{If a line connecting a small degree node is overloaded, it results in an outage in a particular locality only. Hence the blackout is localized and does not affect the remote connections. On the other hand, if higher degree nodes, which are most often generators, get disconnected, then redistribution of power flow causes overload in remote connections, also leading to sequential failures.} 
Such cascading failures have various possible sources. A few of them are listed below:

\begin{itemize}

\item The redistribution of the load on a node or a link which leads to overload\cite{motter2002cascade} or under-load

\item The existence of direct dependency, where if a node becomes dysfunctional, all nodes that are dependent on it also become dysfunctional \cite{parshani2011critical}

\item The number of functioning adjacent nodes is above or below threshold\cite{baxter2010bootstrap}.
\end{itemize}

There are several models proposed for the study of cascading failure using simulations in AC power grids, like link overload-based models as well as node failure-based models. In the node failure model reported \cite{wang2009cascade}, the tolerance parameter $T$ is calculated for each node, such that if it goes beyond a critical threshold $T_C$, the node is considered to be dysfunctional. The link overload-based model reported recently\cite{motter2002cascade}, assumes that the power is exchanged along the shortest path between any two nodes of the given network. Since the power flow from one node to another one does not always take place along the shortest paths, in \cite{CFJurgen}, an improved version of this model is introduced, which is the method followed in the present study. We briefly discuss below the steps involved in analyzing cascading failure using this model and apply them to the synthetic network and the Indian power grid.

In this model, the input power at each node of the grid is taken using the following equation for active power $P_i$\cite{CFJurgen}.

\begin{equation}
    P_i = \sum_j a_{ij} \frac{|V_i||V_j|}{x_{ij}} \sin \theta_{ij}
    \label{eq:powernonde}
\end{equation}

\noindent Here, $a_{ij}$ is the element of the adjacency matrix, $\theta_{ij}$ is the angular difference between the voltages of two nodes $V_i$ and  $V_j$. $x_{ij} = x'. d_{spatial}(i,j)$ is reactance with $x'$ being specific reactance and $d_{spatial}(i,j)$, the length of the transmission line. We consider the value of the specific reactance to be $x'=0.265 \Omega/km$\cite{PSDbook}. Since we are considering high-voltage AC grids, resistance in transmission lines can be conveniently neglected. 

Following this, we calculate the power transmitting through each link of the grid at every time step by solving the nonlinear equation (\ref{eq:powernonde}) to calculate the $\theta_{ij}$. We use it to calculate the power flow in the transmission line connecting nodes $i$ and $j$ as $F_{ij}$ given by

\begin{equation}
    F_{ij} = \frac{|V_i||V_j|}{x_{ij}} \sin \theta_{ij}
    \label{eq:pflow}
\end{equation}

Ideally, total power in the power grid should equate to zero for conservation, i.e.,  $\sum_i P_i = 0$ \cite{CFJurgen}. It is required for the optimum loading on each of the nodes and its stable function. In cases of disturbance, this sum may not be zero at some nodes, which leads to an overload in transmission lines. Up to a certain extent, transmission lines can tolerate the excess power flow without failing. 
The critical power flow that the link can tolerate is given by $C_{ij}$

\begin{equation}
    C_{ij} = (1+ \alpha ) F_{ij}(t_0)
    \label{eq:secureflow}
\end{equation}

\noindent We set the tolerance parameter $\alpha = 0.3$ for every edge in the network in equation (\ref{eq:secureflow}). This would mean that every link can bear the overload up to $1.3$ times the safe power limits at initial time $t_0$.

{We apply this method first to the synthetic exponential network constructed, as mentioned in section III. {From available data, we estimate the number of generators or power distribution stations in the Indian power grid to be approximately $1/5$ of that of the total number of nodes. So, we take $N/5$ high-degree nodes of the network as generators, with the input power positive and the rest as consumers with negative input power.} 
Also, in the initial step, we take $P_i = 1$ for generators and $P_i = -1/4$ for consumers so that the net power in the network is zero.}

We study the cascading failures in the synthetic power grid network with two types of initial failing links. In the first case, the link connecting two high-degree nodes, specifically a generator to a generator, is failed. In the other case, we choose a link that fails as the one connecting a dead-end node to the power grid. At every step, to make the net power equal to zero, we regulate the power at each node using the equation
{
\begin{equation}
    P_i' = P_i - \frac{1}{N(\zeta)} \sum_{j \in \zeta} P_j
    \label{eq:pwrreg}
\end{equation}

\noindent where $N(\zeta)$ is number of nodes in component $\zeta$.}  Then we calculate the flow in each link and the link for which $F_{ij}>C_{ij}$ is removed. We repeat the same procedure until there is no link satisfying this condition.

\begin{figure}[ht]
    \centering
    \includegraphics[scale=0.85]{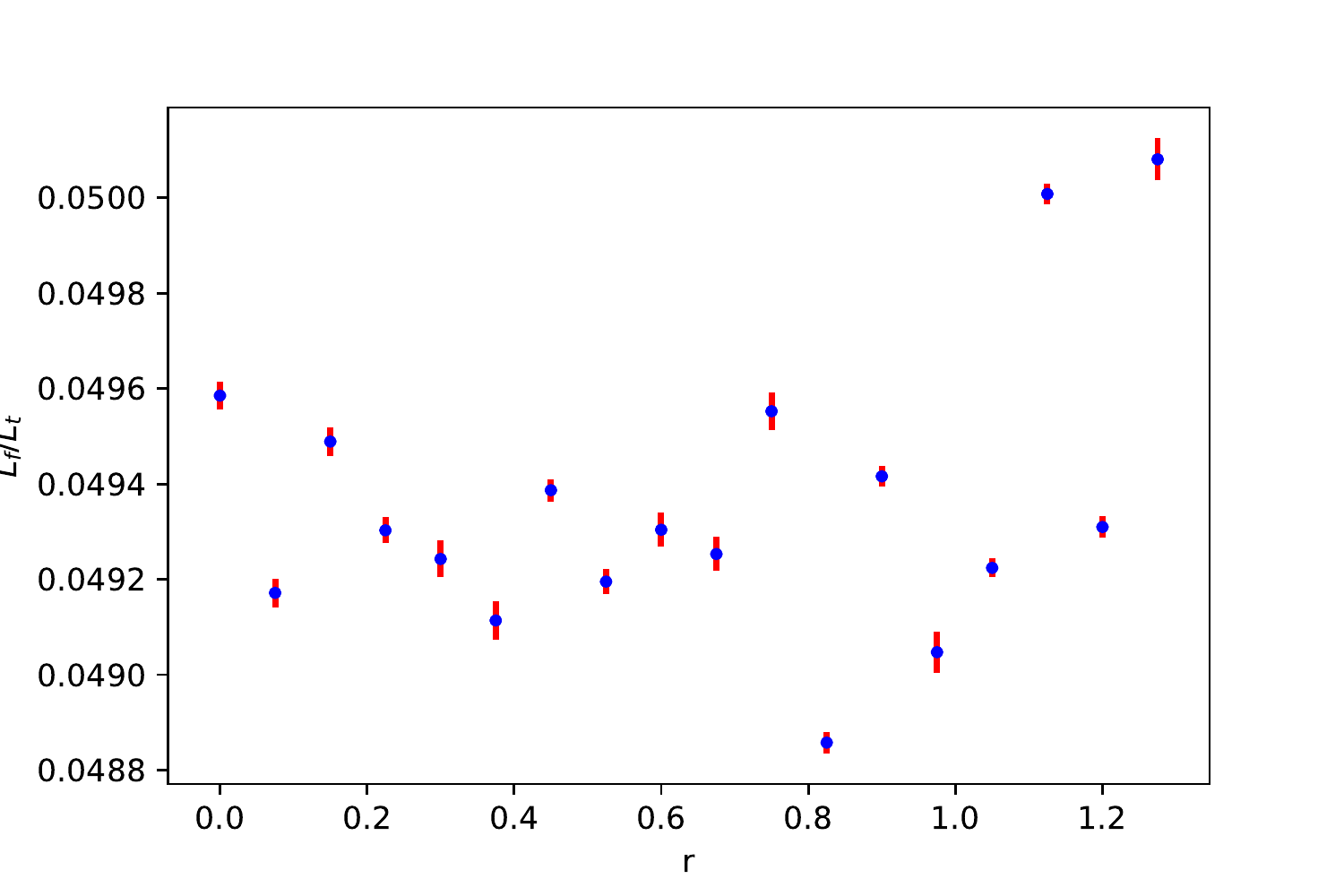}
    \caption{{Variation in the extent of cascading failure in exponential networks as a function of redundancy parameter $r$. Here the mean values of the ratio of failed links, $L_f/L_t$, are shown for 30 realizations of the constructed network with $s = 0.3$, with standard deviations as error bars(in red).}}
    \label{Fig:efailVSr}
\end{figure}

\begin{figure}[ht]
    \centering
    \includegraphics[scale=0.85]{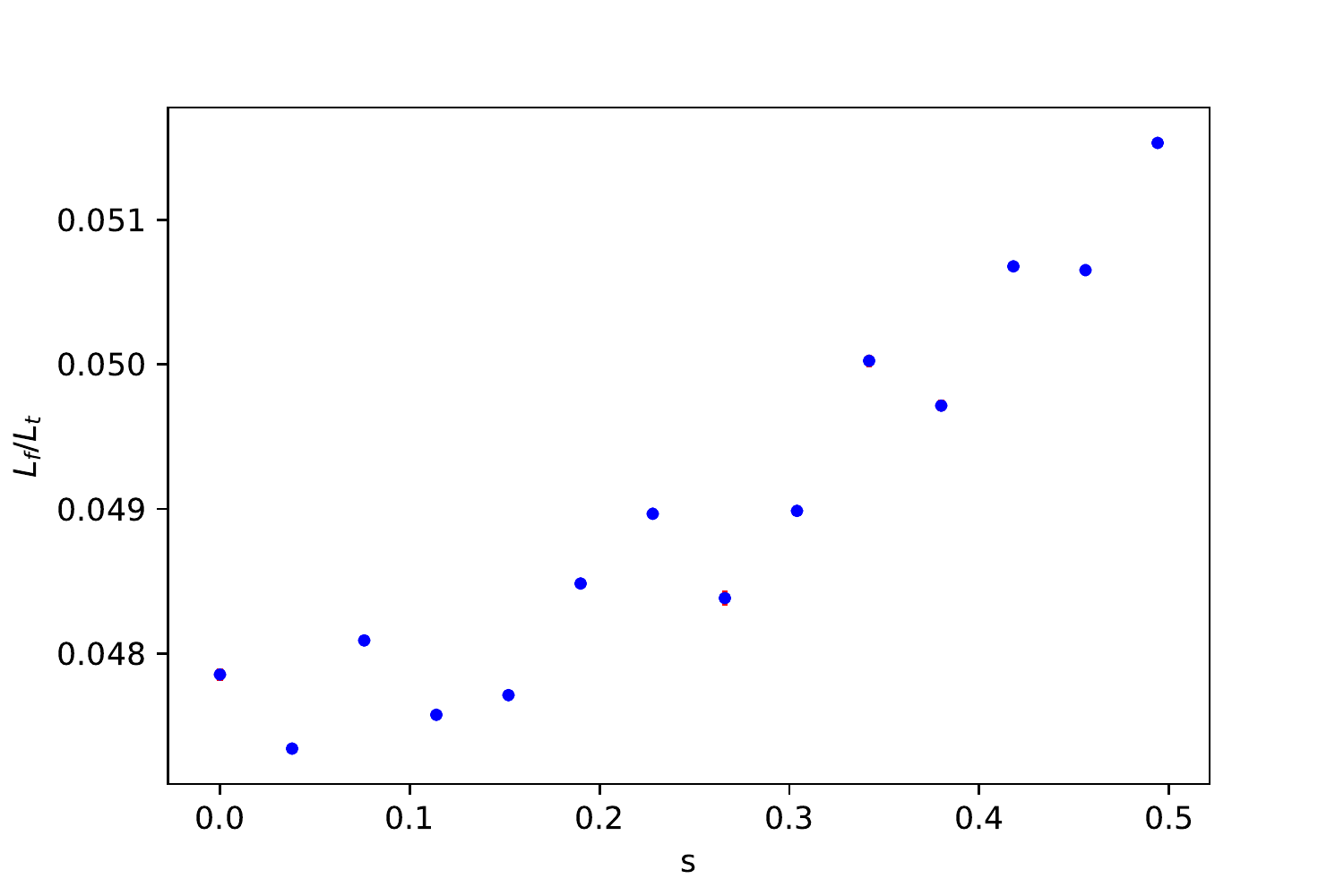}
    \caption{{Variation in the extent of cascading failure in exponential networks as a function of dead-end parameter $s$ keeping $r = 0.1$. The mean values of the ratio of failed links, $L_f/L_t$ from 30 realizations of the constructed network are plotted with standard deviations as error bars(in red).}}
    \label{Fig:efailVSs}
\end{figure}

We choose the parameters of synthetically constructed exponential networks as $(N, N_0, p, q) = (655, 100, 0.3, 0.2)$, equal to the estimated ones for the Indian power grid. 
The distances $d_{spatial}(i,j)$ are taken to be of the same order of magnitude as the Indian power grid and the load $|V_i|=|V_j|=400kV$.
We vary $r \in [0,1.2]$ and $s \in [0,0.5]$ and study the dependence of the spread of failure on the topology of the network. From $30$ realizations, we compute the ratio of failed links to total links, $L_f/L_t$, that will give a measure of the extent of the spread of failure over time.

{When a link connecting a generator to a generator is chosen to initiate the failure, the values of $(L_f/L_t)$ with change in parameter $r$, keeping $s = 0.3$ are shown in Figure \ref{Fig:efailVSr}. We observe that this ratio first decreases up to a certain value and then increases, indicating a minimum value at $r=0.8$. This is because when $r$ is  small, with less number of redundant connections, the excessive power cannot be redistributed among its links, which results in the failure of many links. On the other hand, on networks with large $r$ and having too many redundant connections, the failure spreads across the network. In between, there can be a value of $r$ for the given network at which $L_f/L_t$ can be small, indicating less damage to the network.

In the same manner, we study the effect of parameter $s$ on the failure ratio of links keeping the value of $r$ fixed to $0.38$. As shown in Figure \ref{Fig:efailVSs}, it has an increasing trend with parameter $s$. Since large $s$ corresponds to less number of dead-ends, it is clear that the failure cannot be contained, and the number of failed links increases.

However, when a link connecting dead-end to the main network is set to fail initially, a maximum of two links and a few nodes and links that are directly dependent on it are isolated from the main network, leaving the remaining network unaffected.
We note that the values of $r$ and $s$ that give a minimum $(L_f/L_t)$ may also depend on the size of the network, chosen values of $N_0$, tolerance parameter $\alpha$ and the initial failing link.
Our results indicate that for a given power grid, an optimum redundancy and dead-ends can be found to minimize the extent of cascading failure.}

%-----------------------------------------------------
%---------------    Section IV    --------------------
%-----------------------------------------------------
\section{Cascading failure on the Indian power grid}

Once the stability of the synthetic exponential networks is studied and quantified, we apply the same link failure model to the Indian power grid to analyze its robustness. We use the networks of 220kV lines and 400kV lines and apply the same method described in the previous section. Since the data on lengths of transmission lines is not available for the North-Eastern regional power grid, that is not included here. 

We choose the tolerance parameter $\alpha = 0.3$ and randomly choose $N/5$ nodes with higher degrees to be generators. 
We take $|V_i|=|V_j|=220kV$ and $|V_i|=|V_j|=400kV$ for $220kV$ and $400kV$ networks respectivly.
We simulate cascading failures as described in the previous section, and the extent of link failures are shown in Figures \ref{fig:cf220} and \ref{fig:cf400} for the $220kV$ and $400kV$ networks. Here failed links are shown in {red dotted lines} on the network with active links in green.
{For the $220kV$ grid, we find that $L_f/L_t = 0.045$. For the $400kV$ grid, we calculate this ratio to be $0.038$, making it slightly more robust than the $220kV$ network. 
The spread of the cascade on the network in time is compared for both these networks in Figure \ref{fig:cfprog} where the changes in $L_f/L_t$ are plotted against time steps. The ratio saturates in time as the spread of overloading stops.
We note the ratio is higher but saturates earlier for the 220kV grid compared to the 400kV grid. 
Thus, the 400kV grid fails slowly but to a lesser extent compared to the 220kV grid. 
In both these cases, we note that evenly distributed generators help these networks to separate into two stable and functioning clusters, each having generators and consumers. However, in this process, over 200 nodes and associated links are isolated from the two main clusters, and the cascade terminates.

We carry out a detailed analysis to identify links that can cause an extensive spread of cascades. For this, we repeat the same analysis by considering each link in each network as the starting link to fail and compute the corresponding $L_f/L_t$ values. 
{We find the links connecting to dead ends, with one end being a node of degree $k_i =1$, are non-critical.} 
The $L_f/L_t$ value is minimum for each of them and the cascade does not spread in the network. In $220kV$ network, we find $15\%$ links are of this type, which leads to the value of $L_f/L_t =0.002$. In the $400kV$ network, we find $21\%$ links connecting dead ends that give $L_f/L_t= 0.001$. On the other hand, links connecting high-degree nodes, with degrees more than or equal to $5$ on either or both sides, are critical and they most often connect generator to generator.
In the 220kV network, there are $5\%$ links that give $L_f/L_t$ values in the range of $0.039$ to $0.055$, and $2\%$ links in $400kV$ with $L_f/L_t$ in the range $0.025$ to $0.035$.
We also find links with tree-like structures on either or both sides that can be identified by the local clustering coefficients of their connecting nodes. In $220kV$ network we have $63\%$ of such links, with $L_f/L_t$ in $(0.013, 0.035)$ and in $400kV$ network, there are $50\%$ of such links with $L_f/L_t$ in $(0.006, 0.025)$. 
However, there are also links that connect nodes, with degrees equal to $2$ or $3$ on either side but result in higher values of $L_f/L_t$. 
We find they lie on the shortest paths between two generators. The remaining links mostly have clustered components on both sides and, depending on the extent of clustering, can result in a range of $L_f/L_t$ values, as $0.01–0.039 (220kV)$ and $0.004-0.025 (400kV)$

\begin{figure}[ht]
    \centering
    \includegraphics[scale=0.55]{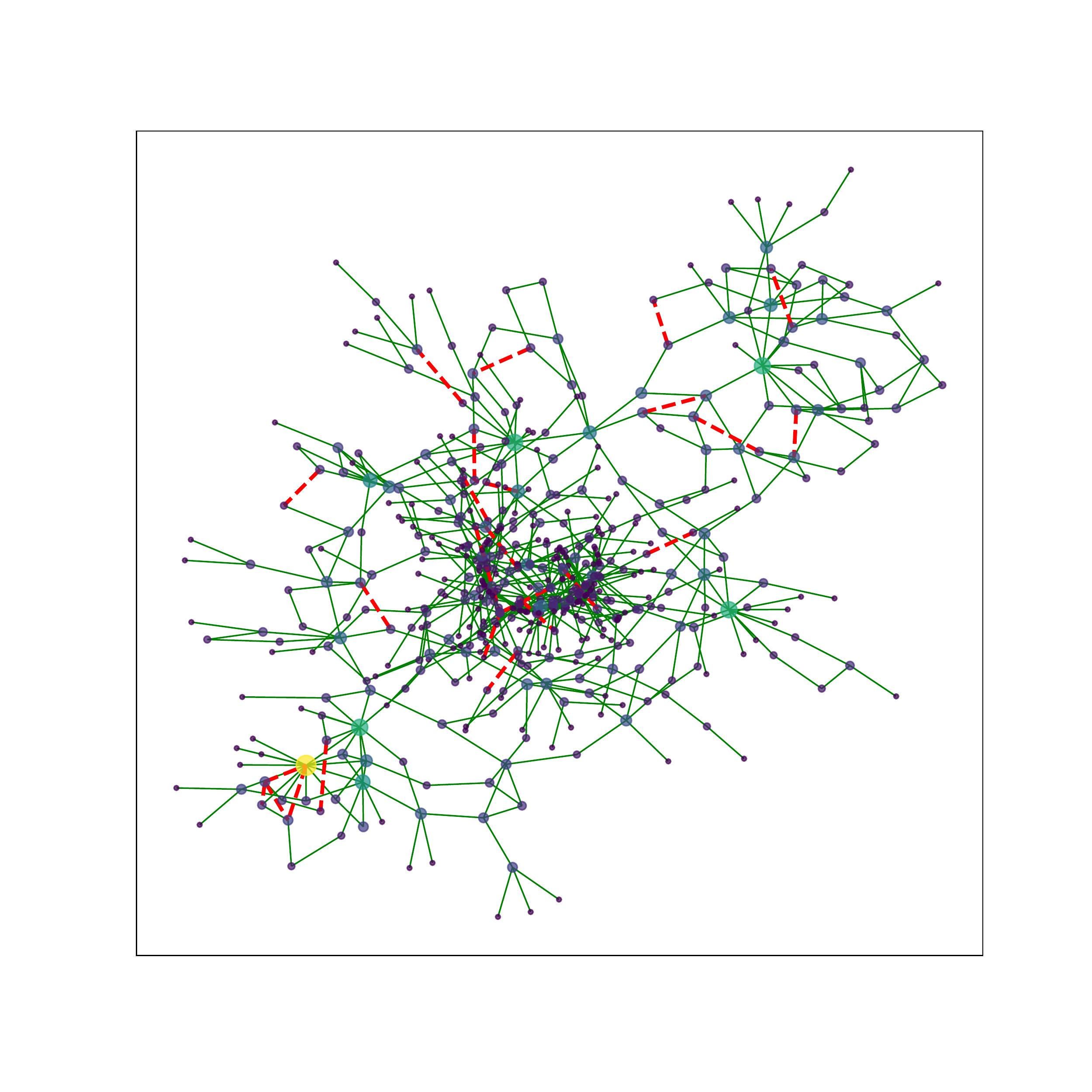}
    \caption{{Cascading failure in the 220kV transmission network of the Indian power grid, where 0.045 of the total links are failed (shown in red dotted line)}}
    \label{fig:cf220}
\end{figure}

\begin{figure}[ht]
    \centering
    \includegraphics[scale=0.55]{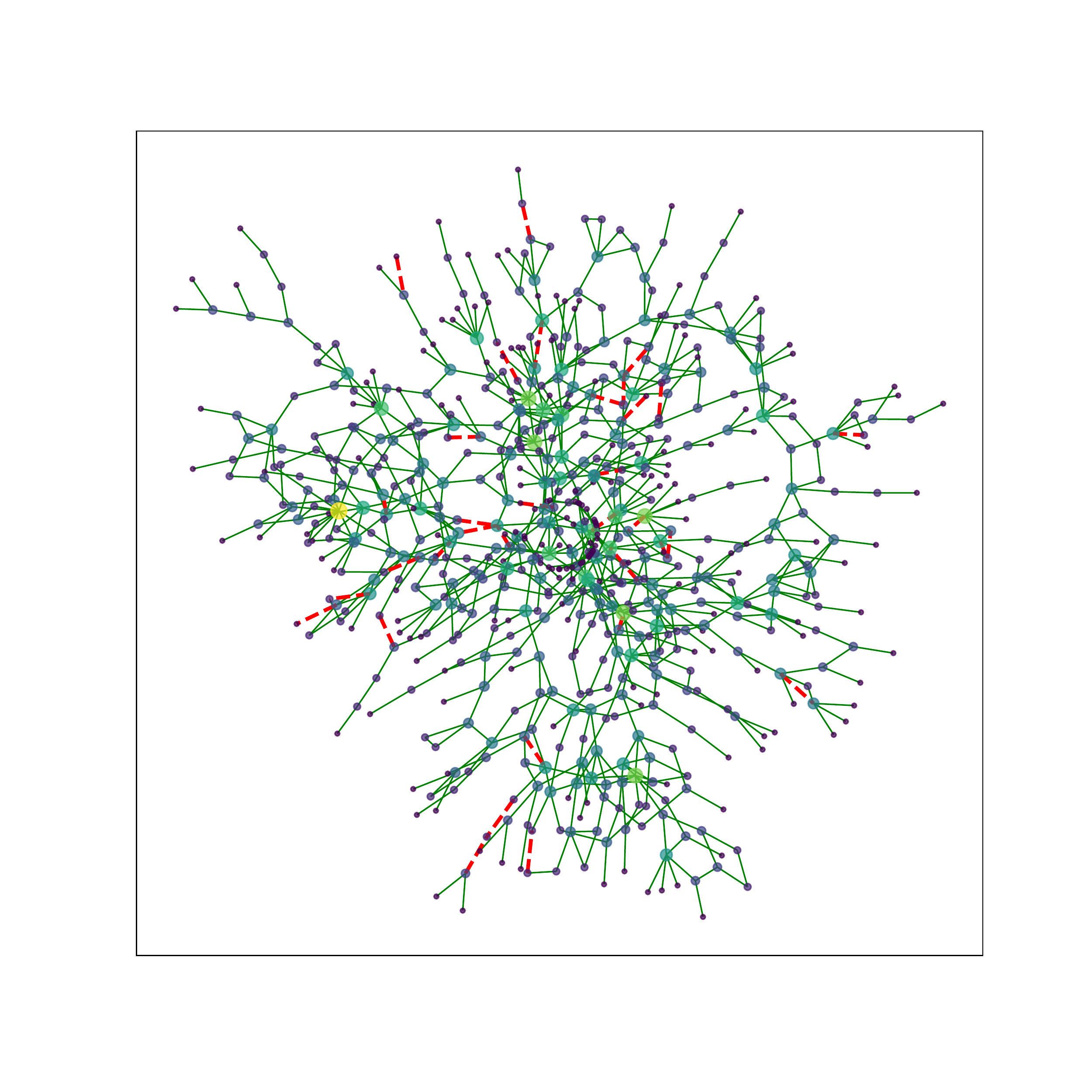}
    \caption{{Cascading failure in the 400kV transmission network of the Indian power grid. Here the failed links(shown in red dotted line) are 0.038 of the total links.}}
    \label{fig:cf400}
\end{figure}

\begin{figure}[]
    \centering
    \includegraphics[scale=0.65]{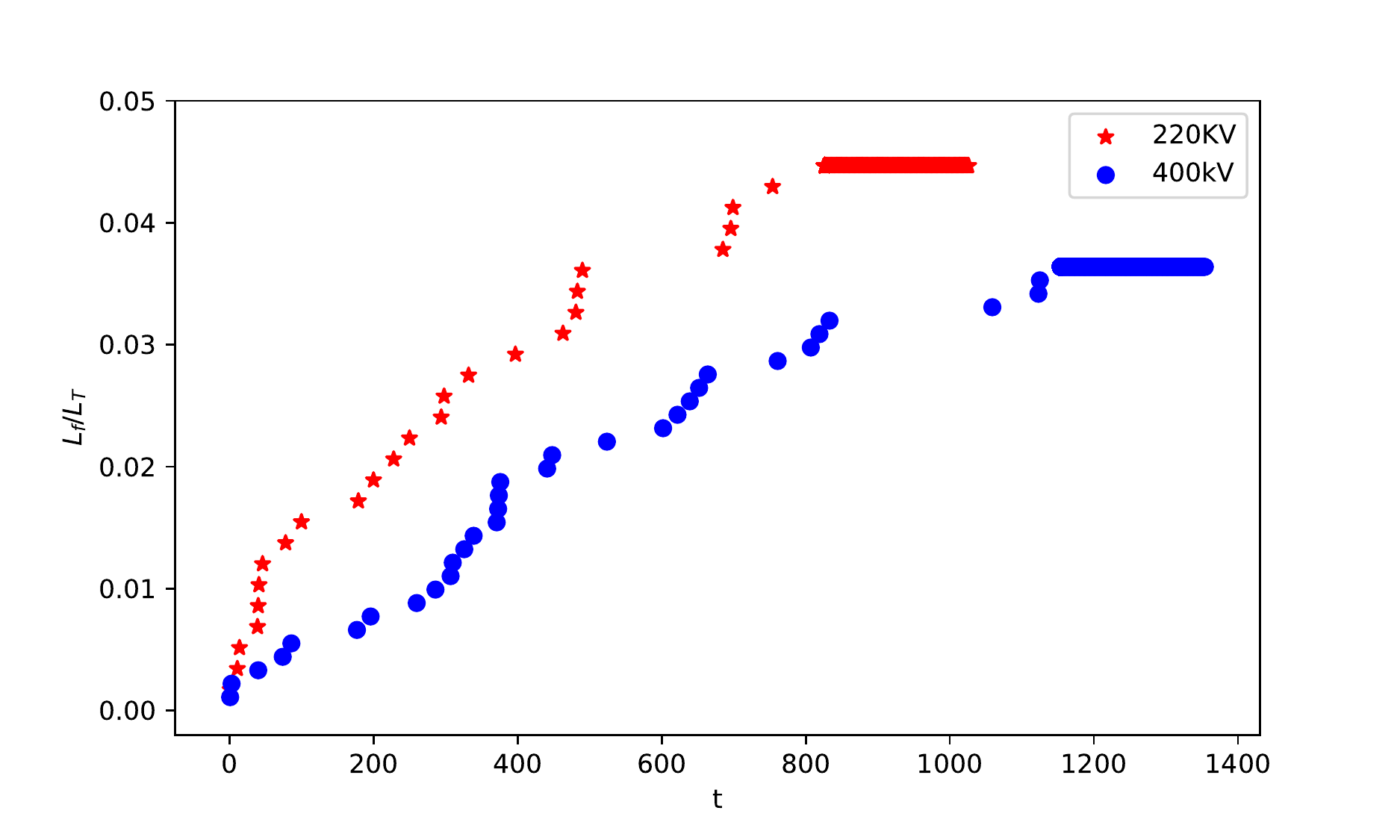}
    \caption{{Variation of the link failure ratio, $L_f/L_t$ with time in 220kV network and 400kV network. At any time step, the extent of failure in the 400kV network is less compared to the other one, but it takes more time to reach saturation.}}
    \label{fig:cfprog}
\end{figure}

%-----------------------------------------------------
%----------------    Section V    --------------------
%-----------------------------------------------------
\section{Conclusion}

From the data available with POSOCO(as on July 2021), the power transmission network of India is constructed with 220kV and 400kV transmission lines and its structural characteristics are computed. We find that the Indian power grid network is exponential in nature with scaling index $\gamma = 1.78$, similar to the Italian power grid($\gamma = 1.8$) and Western US grid ($\gamma = 2.21$). 

To understand the efficient and stable functioning of the power grids, it is important to calculate the dependence on the features of the network, like redundant connections and dead-ends. For this, we construct a synthetic exponential network that is topologically equivalent to the Indian power grid and report how the network measures change with redundancy and dead-ends. {With increasing redundancy, $CC$ of the network decreases while $GE$ shows an increasing trend. With an increase in $s$ or a decrease in the number of dead-ends in the network, $CC$, $GE$ and $\bar{k}$ exhibit decreasing trends. These trends in the network measures are important to understand the dependence of the cascading failures in the structure of power grid networks.}

We study the spread of cascading failure due to the failure of a link connecting one generator to the other in the synthetic exponential network and its robustness as network features are changed. {We find that for a given size of the network, it is possible to obtain optimum values for parameters related to redundancy and dead-ends that can minimize the failure ratio $(L_f/L_t)$ and hence provide maximum resilience against cascades. On the contrary, if a link connecting a dead-end with the main cluster fails, it  leads to a local blackout only without affecting the stability of the remaining network.}

We apply the link failure model to study cascading failures in the Indian power grid. We find the structure of the Indian power grid is such that $4.5\%$ of links in the $220kV$ network and $3.8\%$ of links in the $400kV$ network are seen to fail when $N/5$ nodes are taken as generators. Moreover, the networks form two stable and functioning clusters.

{We find the spread of the cascade is maximum when the link that fails first is the one connecting generator to a generator in the network. The spread is minimum when it starts with links that connect to dead ends. The extent of spread also depends on whether the starting links are the ones that connect to tree like structures on either or both ends or are connected to clustered components.}

Our study on the synthetically generated exponential network and the Indian Power grid brings out how the spread of failures depends on the network structure and topological features. This can guide engineers to choose a proper tolerance parameter to avoid the probability of small-scale and large-scale cascading failures and also to arrive at the most efficient way of adding nodes or extending connections for optimum performance.

\section{Data Availability}
The data used is obtained from POSOCO’s individual regional load dispatch centers (RLDCs)(https://posoco.in) as on July 2021, for {Eastern}, Northern, Western and Southern regions. {For North Eastern region, the data was collected from the Central Electricity Authority, Ministry of Power, Government of India(https://cea.nic.in/)} 

\subsection*{Acknowledgements}
{One of the authors(VT) acknowledges the Department of Science and Technology(DST) India for INSPIRE scholarship and IISER Tirupati for facilities during MS thesis work.}
\appendix

\subsection*{References}
%\nocite{*}
% Produces the bibliography via BibTeX.
%\printbibliography

\end{document}